\documentclass[preprint,preprintnumbers,amsmath,amssymb]{revtex4-2}
\usepackage{amsfonts}    
\usepackage{amssymb}
\usepackage{latexsym}
\usepackage{graphicx}
\usepackage{rotating}
\usepackage{multirow}
\usepackage[usenames,dvipsnames]{color}
\usepackage[active]{srcltx}
\usepackage[colorlinks=true]{hyperref}
\usepackage{color}

\newcommand{\al}{\alpha}

\newcommand{\ep}{\epsilon}
\newcommand{\Si}{\Sigma}

\newcommand{\de}{\delta}

\newcommand{\De}{\Delta}

\newcommand{\rar}{\rightarrow}
\newcommand{\lrar}{\leftrightarrow}
\newcommand{\non}{\nonumber}

\newcommand{\re}[1]{(\ref{#1})}
\newcommand{\phm}{\phantom{-}}
\begin{document}

\title{Towards the analytic theory of Potential Energy Curves for diatomic molecules. Studying He${}_2^+$ and LiH diatomics as illustration.\\
}
\date{\today}

\author{Alexander~V.~Turbiner}
\email{turbiner@nucleares.unam.mx}
\affiliation{Instituto de Ciencias Nucleares, Universidad Nacional
Aut\'onoma de M\'exico, Apartado Postal 70-543, 04510 M\'exico,
D.F., Mexico}
\author{Horacio~Olivares-Pil\'on}
\email{horop@xanum.uam.mx}
\affiliation{Departamento de F\'isica, Universidad Aut\'onoma Metropolitana-Iztapalapa,
Apartado Postal 55-534, 09340 M\'exico, D.F., Mexico}

\begin{abstract}
Following the first principles the elements of the analytic theory of potential
curves for diatomic molecules (diatomics) are presented. It is based on matching
the perturbation theory at small internuclear distances $R$ and multipole
expansion at large distances, modified by exponentially small terms for
homonuclear case, with addition of the phenomenologically described
equilibrium configuration, if exists. It leads to a new class of (generalized)
rational potentials (modified by exponentially small terms) with difference
in degrees of polynomials in numerator and denominator equal to 4 (6)
for positively charged (neutral) diatomics.

As examples the He$_2^+$ and LiH diatomics in Born-Oppenheimer approximation
are considered. For ${}^{4}$He$_2^+$ (${}^{3}$He$_2^+$) diatomics it is found
the approximate analytic expression for the potential energy curves (analytic PEC)
$V(R)$ for the ground state $X^2 \Si_u^+$ and the first excited state $A^2 \Si_g^+$.
It provides 3-4 s.d. correctly for distances $R \in [1, 10]$\,a.u. with some
irregularities for $A^2 \Si_g^+$ PEC at small distances
(much are smaller than equilibrium distances) probably related to level
(quasi)-crossings which may occur there.
The analytic PEC for the ground state $X^2 \Si_u^+$ supports 829 (626) rovibrational states
with 3-4 s.d. of accuracy in energy, which is only by 1 state less (more) than 830 (625)
reported in the literature. In turn, the analytic PEC for the excited state $A^2 \Si_g^+$
supports all 9 reported weakly-bound rovibrational states.
For ${}^{7}$LiH it is found the analytic expression for the ground state $X^1\Si^+$ PEC
in the form of rational function, which supports 906 rovibrational states with
3-4 s.d. accuracy in energy, it is only of 5 states more than reported in the literature (901).
For both diatomics the difference in number of rovibrational states is related with
the non-existence/existence of weakly-bound states close to threshold (to dissociation limit).
Entire rovibrational spectra is found in a single calculation using the code based on
the Lagrange mesh method.
\end{abstract}

\maketitle
\section{Introduction}

Developed in molecular physics the celebrated Born-Oppenheimer (BO) approximation \cite{BO}
(for discussion see \cite{LL}) takes advantage of the large difference in the masses of the
nuclei and electrons. This approach allows to separate the fast (electronic) degrees of freedom
from the slow (nuclear) ones. At the first stage in this approximation the nuclei are assumed to be clamped
forming a certain spatial configuration, thus, nuclear masses are infinite and internuclear distances
become classical, non-dynamical variables. Sometimes, in physics literature it is called
the Born-Oppenheimer approximation of the zero order.
Hence, the original many-body Coulomb problem becomes the problem of electrons interacting with several fixed Coulomb (charged) centers. Its dynamics is governed by the electronic Hamiltonian,
\begin{equation}
\label{Hel}
  {\cal H}_{e} \ =\ T_e\ +\ V(r_e; R_N)\ ,\
\end{equation}
where the electronic kinetic energy
\[
   T_e\ =\ -\frac{1}{2}\sum_i \nabla^2_i (r_e)\ ,
\]
is sum of kinetic energies of individual electrons, where $\hbar=1, m_e=1$, and the Coulomb potential energy, which is the sum of three terms,
\[
   V(r_e; R_N)\ =\ V_n(R_N)\ +\ V_{en}(r_e, R_N)\ +\ V_e(r_e)\ .
\]
These terms correspond to internuclear interaction, electron-nuclear interaction and electron-electron interaction, respectively. The coordinates of the nuclei $\{R_N\}$ are fixed becoming classical (non-dynamical), they play a role of the external parameters in the Hamiltonian (\ref{Hel}). Due to translation-invariant nature of the Coulomb potential $V(r_e; R_N)$ the relative distances between particles appear in the potential only. One can calculate in a straightforward way, usually, numerically, the spectra of the electronic Hamiltonian ${\cal H}_{e}$ getting the spectra of the electronic energy $E_e$, in other words, the electronic terms. The obtained electronic energy, or, in other words, the eigenvalue of the electronic Hamiltonian ${\cal H}_{e}$, depends on the nuclear configuration, thus, leading to the so-called Potential Energy Surface (PES) $V(R_N)$, where the nuclear configuration occurs as the ``argument". It implies that the internuclear distances $R_{ij}=|{\bf R}_i-{\bf R}_j|$ are positive, they are {\it not} dynamical variables. In the second stage the internuclear distances $R_{ij}$ are restored as dynamical variables and PES $V(R_N=R_{ij})$ is taken as the potential for nuclear motion assuming its nuclear mass independence. It is worth noting that in this approach due to smallness the mass of electrons $m_e$ in comparison with the mass of nuclei $m_N$ the center-of-mass of the original many-body Coulomb system differs {\it slightly} from the center-of-mass of nuclei. It does not affect PES but it leads to some small corrections of the order of $m_e/m_N$ in study of nuclear motion (for discussion, see \cite{Cederbaum:2013} and references therein). They will be neglected in the present study, since we are focused on the properties of PES.

It must be emphasized that for polyatomic molecules the PES $V(R_N)$ can be presented as the sum over 2-,3- etc body interactions. Schematically, it can be written as
\[
  V(R_N) \ =\ V_{2-body}(R_{12})+V_{2-body}(R_{13})+V_{2-body}(R_{23})+\ldots\ +\ V_{3-body}(R_{12},R_{13},R_{23})\ +\ \ldots \quad .
\]
It must be emphasized that 3- and more body interactions are of short range: they die out for large internuclear distances, see \cite{Teller:1943} and numerous references into there.
In the present paper we will be focussed on diatomic systems ($N=2$) where $V(R_{N=2})=V(R_{12})$,
where $R_{12}$ is the relative distance between the two nuclei (see below), since now on
we denote $R \equiv R_{12}$.

\begin{figure}[h!]
\begin{center}
\includegraphics[width=7cm]{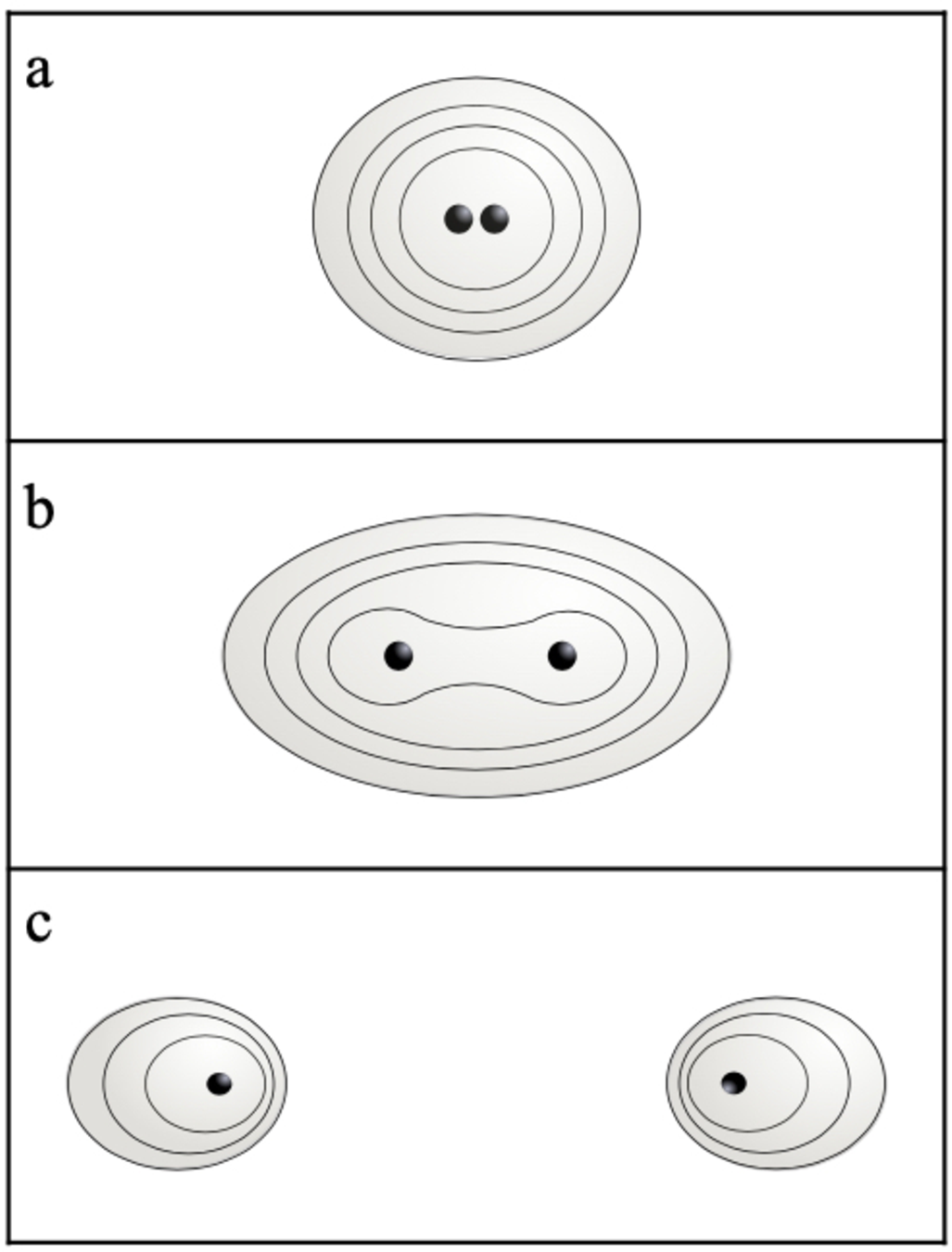}
\caption{Schematic representation of the electronic clouds for a diatomic molecule for ($a$) small,
($b$) intermediate and ($c$) large internuclear distances. Nuclei are marked by bullets.}
\label{figMol}
\end{center}
\end{figure}

In physically important case of diatomic molecules and ions, in particular, of singly-positively-charged diatomic molecular ions of the type $(A + B)^+$ with nuclear charges $Z_{A}$ and $Z_{B}$
with $(Z_A+Z_B-1)$ electrons and neutral diatomic molecules $(A + B)$ with $(Z_A+Z_B)$ electrons,
on which we will be focused, the nuclear configuration is defined by the single internuclear
(classical) distance $R > 0$. In this case the PES becomes the Potential Energy Curve (PEC)
$E_e=V(R)$. From physics point of view, taking charges $Z_{A,B}$ as {\it probes}, the potential
\[
V(R)=\frac{Z_{A} Z_{B}}{R}\,S(R)\ ,
\]
measures the screening $S(R)$ of Coulomb interaction of nuclei due to the presence of electronic media
In this consideration it is always assumed that the nuclei are point-like,
having no internal structure. We never go to ultra-small distances where strong interaction
plays the role and nuclear forces should be considered,
see for illustration Fig.\ref{figMol}.
Sometimes, the interplay between the Coulomb repulsion at small distances and the Van-der-Waals attraction
at large distances leads to sufficiently deep minimum in $V(R)$ at distances of order of 1 a.u. (other than Van-der-Waals minimum which is always present and situated usually at large internuclear distances
of the order of 10 a.u.) manifesting the existence of the bound states and, finally, the molecule.
In the rotation frame the potential $V(R)$ is accompanied by centrifugal term.
These states are called the rovibrational states.

Since early days of quantum mechanics the potential curves attracted a lot of attention while
the main emphasis was given to the creation of the models which describe the vicinity of the
potential well in maximally accurate manner: the harmonic and Morse oscillators,
P\"oschl-Teller, Lennard-Jones and Buckingham potentials, and their numerous modifications,
see e.g. \cite{Yanar:2016} and references therein and into as well as a recent review \cite{AB:2021}.
Usually, these models were pure phenomenological, they never pretend to describe the whole potential
curve and total rovibrational spectra other than several lowest quantum states situated close
to minimum of the PEC and the so-called anharmonicity constants. However, since long ago
the present authors considered as the challenge to find a function, which would describe
the {\it ab initio} calculations carried out by Kolos and Wolniewicz \cite{Kolos-Wolniewicz:1965}
for the PEC of the hydrogen molecule H$_2$ as well as experimental data adequately.

A new development happened in \cite{OT:2018} where for the first time for three simple molecules
H$_2^+$, H$_2$, (HeH) a certain (generalized) meromorphic (rational) potentials of a new type
were proposed which admitted to find the whole rovibrational spectra for a set of electronic terms
(PEC) and even some transition amplitudes between different electronic terms \cite{OT:2016}.
Those potentials were characterized by the correct asymptotic behavior at small and large
internuclear distances. In concrete, the potentials had a form of (generalized) rational
functions modified by exponential terms multiplied by other rational functions
for the case of homonuclear diatomic molecules.

The goal of this paper is to develop systematically a theory and construct simple analytic
potentials modelling PEC which are able to describe the rovibrational spectra of diatomics.
Note that these potentials result from the interpolation between small and large distance
behavior of PEC. In general, they do not assume {\it a priori} the existence of the global
minimum at $R \sim 1$\,a.u.: it occurs naturally due to concrete, physics-inspired coefficients
in perturbation theory and multipole expansion. The the minimum can be tuned phenomenologically
to describe the experimental data. As an illustration of the general construction,
two diatomic molecules He$_2^+$ and $^7$LiH will be considered. Since now on in order
to simplify notations sometimes we will address to diatomic molecules and diatomic
ions as {\it diatomics}.

Atomic units are used throughout the paper although the energy is given in Rydbergs.

\section{Generalities}

Making analysis of the electronic Hamiltonian for $(A + B)^+$ ion one can find that the
potential $V(R)$ at small $R$ is defined via perturbation theory in $R$,
\begin{equation}
\label{VRsmall}
V(R)\ =\ \frac{Z_{A} Z_{B}}{R}\ +\ E_{u.i.}\ +\ E_1\, R + O(R^2)  \ ,
\end{equation}
where the first (classical) term comes from the Coulomb repulsion of nuclei, the second term
$E_{u.i.}$ is defined by the energy of united ion (u.i.) with total nuclear charge
$(Z_{A} + Z_{B})$ with $(Z_{A} + Z_{B}-1)$ electrons, see e.g. \cite{LL}, $\S$ 80;
sometimes, it is called the {\it super-ion}. It was observed that the
linear term $\sim R$ is always absent, $E_1=0$, at least, for the ground state \cite{RAB:1958,WAB:1959,OT:2018}.

At large distances $R$ for the potential curve the leading term of the interaction
of neutral atom with charged atomic ion is given by Van-der-Waals attraction potential
with higher corrections in powers of $1/R$,
\begin{equation}
\label{VRlarge}
   V(R)\ =\ -\frac{c_4}{R^{4}}\ +\ \frac{c_5}{R^{5}}\ +\ \frac{c_6}{R^{6}}\ +\
   \frac{c_7}{R^{7}} \ +\ \ldots  \ ,\ c_4 > 0\ ,
\end{equation}
see \cite{MK:1971} and for recent (extended) discussion, \cite{IK:2006}. Here the parameters
$c_{4, \ldots}$ are related with (hyper)polarizabilities of different orders, see \cite{LL}, $\S$ 89.
For the case of the ground state $c_5=0$ and, hence, the term $\sim \frac{1}{R^{5}}$ in (\ref{VRlarge})
is absent ~\cite{MK:1971,IK:2006}. The first term $\sim 1/R^4$ in (\ref{VRlarge}) has a meaning of the charge-induced dipole interaction potential, the term $\sim 1/R^6$ corresponds, basically,
to charge-induced quadrupole interaction potential.

For the case of neutral diatomics $(A + B)$ (in fact, we consider the interaction of two neutral atoms
at large internuclear distances) the first two coefficients in (\ref{VRlarge}) are usually absent,
$c_4=c_5=0$, and we arrive at the expansion
\begin{equation}
\label{VRlarge-atom}
V(R)\ =\ -\ \frac{d_6}{R^{6}}\ +\ \frac{d_7}{R^{7}} \ +\ \frac{d_8}{R^{8}} \ +\ \ldots  \ ,\ d_6 > 0\ .
\end{equation}
If the ground state is considered $d_7=0$, the first term $\sim 1/R^6$ in (\ref{VRlarge-atom})
has a meaning of the interaction potential of two induced dipoles, the term $\sim 1/R^8$ corresponds
to induced dipole-induced quadrupole interaction potential etc,
see e.g. \cite{Pauling:1935} for the case of hydrogen molecule H$_2$.

It is evident that the attraction at large distances together with repulsion at small distances
implies the existence of the minimum of the potential curve for both diatomic molecular ion and molecule.
This minimum occurs at the moment when repulsion is changed to attraction. If this minimum is situated
at large distances and (very) shallow, such a minimum is usually called the Van-der-Waals minimum.
It is worth noting that for the case of the molecular ion $(A + B)^+$ the expansion (\ref{VRlarge})
remains the same functionally for both dissociation channels: $A^+ + B$ and $A + B^+$,
while evidently the expansion (\ref{VRsmall}) at small distances remains the same
for those channels, naturally, it does not see any difference between channels.

In many cases the known potential curves are smooth curves with slight irregularities,
probably, due to effects of the level (quasi)-crossing with other excited states
of the same symmetry, see for discussion e.g. \cite{MK:1971} and \cite{LL}.
It hints to interpolate the expansions (\ref{VRsmall})-(\ref{VRlarge}) using two-point Pade approximation
\begin{equation}
\label{Pade}
 V_{ion}(R)\ =\ \frac{Z_{A} Z_{B}}{R}\ \frac{P_N(R)}{Q_{N+3}(R)}\ \equiv\ \frac{Z_{A} Z_{B}}{R}\ \mbox{Pade}(N / N+3)\ ,
\end{equation}
where $P_N, Q_{N+3}$ are polynomials in $R$ of degrees $N$, $(N+3)$, respectively, with normalization
$P_N(0)= Q_{N+3}(0)=1$, as it was introduced for the first time in \cite{OT:2018} for H$_2^+$ molecular ion, with condition $Q_{N+3} > 0$ for $R>0$. The condition of positivity of denominator $Q_{N+3}(R)$ in (\ref{Pade}) (implying the absence of real positive roots) leads to constraints on its coefficients, not for any $N$ this condition can be fulfilled. This formula seems applicable for {\it any} singly-positively-charged diatomic molecular ion, for both hetero- and homo-nuclear diatomics.
Similar formula can be constructed for the case of neutral diatomics $(A + B)$ with dissociation channel $A + B$ as interpolation of the expansions (\ref{VRsmall})-(\ref{VRlarge-atom}),
\begin{equation}
\label{PadeN}
 V_{neutral}(R)\ =\ \frac{Z_{A} Z_{B}}{R}\ \frac{P_N(R)}{Q_{N+5}(R)}\ \equiv\ \frac{Z_{A} Z_{B}}{R}\ \mbox{Pade}(N / N+5)\ ,
\end{equation}
where $P_N, Q_{N+5}$ are polynomials in $R$ of degrees $N$, $(N+5)$, respectively, with normalization
$P_N(0)= Q_{N+5}(0)=1$ as it was introduced in \cite{OT:2018} for H$_2$ molecule with condition $Q_{N+5} > 0$ for $R>0$. The condition of positivity of denominator $Q_{N+5}(R)$ in (\ref{PadeN}) implies the absence of real positive roots, it leads to constraints on its coefficients, also not for any integer $N$ it can be fulfilled in practice.

Parameter $N$ in (\ref{Pade}) and (\ref{PadeN}) takes integer values, $N=0,1,\ldots$. Polynomials $P_N, Q_{N+k}$ are chosen in such a way that some free parameters in $\mbox{Pade}(N/ N+k)$ for $k=3,5$ are fixed in order to reproduce exactly several leading coefficients in the both expansions (\ref{VRsmall})-(\ref{VRlarge}). Remaining free parameters are found to reproduce the potential at some points in $R$ where the potential is known numerically from solving the electronic Schr\"odinger equation (the so-called {\it ab initio} calculations) or experimentally. It might be considered as the surprising fact that the leading coefficients in the expansions (\ref{VRsmall})-(\ref{VRlarge})-(\ref{VRlarge-atom}) at $R=0, \infty$ {\it ``know"} about the existence/non-existence of the global, non-Van-der-Waals minimum at $R=R_{eq}$. It is worth mentioning if such a minimum exists at $R=R_{eq}$ remaining free parameters in (\ref{Pade}) and (\ref{PadeN}) can be also found by expanding $V_{ion}(R), V_{neutral}(R)$ near minimum making the Taylor expansion around
\[
   V(R)\ =\ -D_e\ +\ \sum_n f_n (R - R_{eq})^n\ ,
\]
where $D_e$ is the dissociation energy, $R_{eq}$ is the equilibrium distance (it corresponds to position of the minimum) and $f_n$ are the molecular or (an)harmonic constants, or, in the Dunham expansion
\[
  V(R)\ =\ D_e \bigg[\left(1 - e^{-\al (R - R_{eq})}\right)^2\ +\ P_3\left(1 - e^{-\al (R - R_{eq})}\right)^3\ +\ P_4\left(1 - e^{-\al (R - R_{eq})}\right)^4\ +\ \ldots \bigg]\ ,
\]
where $\al$ is the Morse constant, $P_{3,4,\ldots}$ are parameters, see \cite{Dunham:1932}.
Important characteristics of the PEC is $R=R_0$, where
\begin{equation}
\label{R0}
  V_{ion}(R_0)\ =\ V_{neutral}(R_0)\ =\ 0\ ,
\end{equation}
and the dissociation energy vanishes. Usually, at $R_0 < R_{eq}$ the potential is positive. For $R > R_0$ the potential gets negative, the bound states (if exist) are localized inside this domain. Influence of the behavior of the potential curve at $R < R_0$ on the bound states positions needs to be investigated.

In the case of identical nuclei $A=B$ (homonuclear case) the system $(A + A)^+$ is
permutationally-invariant $Z_A \lrar Z_B$ and the extra quantum number - parity with respect to interchange of the nuclei positions, situated symmetrically on the molecular axis at $(0,0,-\frac{R_{eq}}{2})$ and $(0,0,\frac{R_{eq}}{2})$ - occurs. This charge exchange energy (or, saying differently, the energy gap) - the difference between two potential curves $\De E=(E_- - E_+)$ of the first ($(k+1)$-th) excited state (of the negative parity) $E_-$ and of the ground ($k$-th excited) state (of the positive parity) $E_+$ - tends to zero exponentially at large $R$,
\[
  \De E\ =\  D_0\, e^{-S_0}(1 + \frac{e}{R} + \ldots)\ ,
\]
where $D_0 > 0$ is monomial in $R$. Furthermore, as the general feature the exponent $S_0 = \al R$ always, where the parameter
$\al$ depends on the diatomics explored \cite{Chib-Jan:1988}, see below.
It implies that these potential curves can be written at large $R$ in the following form,
\begin{equation}
\label{E+- Rlarge}
E_{\mp}\ =\  E_0(R)\ \pm\ \frac{1}{2}\,\de_{\pm}E(R)\ ,
\end{equation}
where $E_0(R)$ is given by multipole expansion (\ref{VRlarge}), it is the same for lowest energy
states of both parities, hence, it does not depends on the state. It is clear that both $\de_{\pm}E$ are
exponentially small. Similar phenomenon of pairing of the states of opposite parities at large $R$ occurs for the potential curves of the $(k+1)$-th and $k$-th excited states.

In the profoundly studied case of H$_2^+$ molecular ion, carried out in the remarkable paper \cite{Cizek:1986}, the expansion
of $\de_{\pm}E$ looks like the so-called {\it trans-series}: it is the expansion in exponentially-small terms (multi-instanton contributions) each of them accompanied by perturbation theory series in $1/R$ of a special structure; it is similar
to the expansion for one-dimensional quartic double-well potential problem,
\begin{equation}
\label{VRlarge-EXP-pm}
\de E_{\pm} =\ D_0 e^{-S_0} \left(1 + \frac{e}{R}\ +\ O\left(\frac{1}{R^{2}}\right) \right)\ \pm \
D_1 e^{-2 S_0} \left(1 + \frac{e_1}{R}\ +\ O\left(\frac{1}{R^{2}}\right) + a \log R \right)\ +\
\ldots  \ ,
\end{equation}
c.f. \cite{Zinn:1981,Dunne-Unsal:2014} (and references therein), where $e, e_1, a$ are
constants, $D_0=\frac{4}{e}R$, $D_1 \sim R^3$. The energy gap has the form
\begin{equation}
\label{VRlarge-EXP}
\De E(R)\ \equiv\ \frac{\de E_- + \de E_+}{2} =\  D_0 e^{-S_0} \left(1 + \frac{e_1}{R}\ +\ O\left(\frac{1}{R^{2}}\right) \right)\ +\ \ldots  \ ,
\end{equation}
where exponent $S_0=R$ looks as the classical action (one-instanton contribution):
tunneling between two identical Coulomb wells, situated at $R=\infty$;
$D_0=\frac{4}{e}R$ looks like the one-instanton determinant in semi-classical tunneling
between two identical Coulomb wells.

In all concrete cases, the present authors are familiar with,
(H$_2^+$, H$_2$, He$_2^+$, Li$_2^+$, Be$_2^+$) diatomic molecular systems the exponent
$S_0$ is linear in $R$ (according to \cite{Chib-Jan:1988}) with coefficients which depends on
the system studied, $S_0=\al R$, see \cite{Chang:1995,OT:2018} and references therein (see Table~\ref{tparmsdE}).
In turn, the factor $D_0$ in (\ref{VRlarge-EXP}) is such that $D_0 \sim R$ for H$_2^+$, $D_0 \sim R^{1/2}$
for He$_2^+$ (see below), $D_0 \sim R^{5/2}$ for H$_2$, see e.g.\cite{OT:2018},
while in all other cases, where it is known, $D_0 \sim R^{\beta}$, hence, it is monomial in $R$
of some degree $\beta$.
It is worth emphasizing that the exponential smallness at large $R$ of the energy gap
implies the well-known fact that the multipole expansions (\ref{VRlarge}) for the ground state and
the first excited state coincide. In turn, at small $R$ the expansion of the energy gap in
$R$ is given by the Taylor series
\begin{equation}
\label{VRsmall-EXP}
\De E(R)\ =\ \de_a\ +\ \de_b R + O(R^2)\ .
\end{equation}
where $\de_a=E_{+}^{u.a}-E_{-}^{u.a.}$ is the difference between
the first excited state and the ground state energies of the united atom (u.a.) in the case of neutral molecule, or
$\de_a=E_{+}^{u.i.}-E_{-}^{u.i.}$ is the difference between
the first excited state and the ground state energies of the united ion in the case of positively-charged molecular ion.

\begin{center}
\begin{table}[!h]
\caption{Parameters $\al, \beta$ in $D_0\sim R^{\beta}, S_0=\al R$ in the energy gap
(\ref{VRlarge-EXP}) for systems: H$_2^+$~\cite{Cizek:1986}, H$_2$~(see \cite{LL}, $\S$ 81, p.315), He$_2^+$~(present work, see Chapter III),  Li$_2^+$ and Be$_2^+$~\cite{Chang:1995,S:2001}.}
\begin{tabular}{c|  lllll}
\\
\hline\hline
           &\ H$_2^+$\ &\ H$_2$\ &\ He$_2^+$\ &\ Li$_2^+$\ &\ Be$_2^+$\\
\hline
    \ $\al$\quad &\ 1\ &\ 2 \  &\ 1.344\  &\ 0.629 \ &\ 0.829  \\
    \ $\beta$ \quad &\ 1  & 5/2 &\ 1/2 & 2.1796 & 1.4125 \\
\hline\hline
\end{tabular}
\label{tparmsdE}
\end{table}
\end{center}
The next step is to construct an analytic approximation of the exchange energy $\De E$
which interpolates the small (\ref{VRsmall-EXP}) and large (\ref{VRlarge-EXP}) internuclear
distances. If $D_0 \sim R^n$ in~\re{VRlarge-EXP}, where $n$ is integer, this is realized
using two-point Pad\'e type approximation
\begin{equation}
\label{Pade-n}
     \De E(R)_{\{n_0,n_{\infty}\}}\ =\ e^{-S_0}\frac{P_{N+n}(R)}{Q_{N}(R)}|_{\{n_0,n_{\infty}\}} \
     \equiv \
      e^{-S_0}{\rm Pade}[N+n/N]_{\{n_0,n_{\infty}\}}(R)\ ,
\end{equation}
where $P_{N+n}(R)$ and $Q_N(R)$  are polynomials of degrees $N+n$ and $N$ respectively.
This approximation supposes to reproduce exactly the first $n_0$ terms at small internuclear distances
and the first $n_{\infty}$ terms at large internuclear distances expansion, respectively.
If degree $n$ in $D_0 \sim R^n$ in~\re{VRlarge-EXP} is half-integer, the formula (\ref{Pade-n}) should be modified,
the change of variable is needed: $r=\sqrt{R}$.
In particular, for $n=5/2$ and $S_0=2R$ (it is the case of H$_2$ molecule), the two-point Pad\'e type approximation
has the form
\begin{equation}
\label{Pade-5/2}
\De E(R=r^2)_{\{n_0,n_{\infty}\}}=  e^{-2\, r^2}\frac{P_{N+5}(r)}{Q_{N}(r)}|_{\{n_0,n_{\infty}\}}
\equiv
e^{-S_0}{\rm Pade}[N+5/N]_{\{n_0,n_{\infty}\}}(r)\ ,
\end{equation}
see \cite{OT:2018}. The case $n=1/2$, which corresponds to the He$_2^+$ molecular ion, will be presented later.
In order to reproduce properly the behavior of the first $n_0$ terms at small internuclear distances \re{VRsmall-EXP} and the first $n_{\infty}$ terms at large \re{VRlarge-EXP} internuclear distances
the evident constraints on the parameters of the polynomials $P_{N+5}(r)$ and $Q_N(r)$ are imposed.
If the degree $\beta$ in $D_0\sim R^{\beta}$ is some real number, see Table~\re{tparmsdE},
a modification of (\ref{Pade-n}) (or (\ref{Pade-5/2})) implies that the regularization
of the determinant (see a discussion above) is needed,
\[
    D_0(R) \rar D_0(R+A)\ ,\ A > 0\ ,
\]
where $A$ is found via fitting the experimental data for the potential curves. It will be done elsewhere.

Due to the exponentially small in $R$ dependence of $\de E_{\pm}$  \re{E+- Rlarge} the main
contribution to the potential curves $E_{\mp}$ at large internuclear distances comes from the mean energy $E_0(R)$,
\begin{equation}
      E_{0}(R) = \frac{E_+\ +\ E_-}{2}\ ,
\end{equation}
see (\ref{E+- Rlarge}).
Neglecting two-instanton contribution ($\sim e^{-2 S_0}$) and other higher order exponentially-small contributions,
the mean energy $E_0(R)$ expansion at large distances is given by~\re{VRlarge}. On the other hand, at small internuclear distances $E_{0}(R)$ expansion has the same structure as~\re{VRsmall} with $Z_A=Z_B \equiv Z$, where
\[
 E_{u.a.}\ =\ \frac{E_{+}^{u.a}+E_{-}^{u.a.}}{2}\ ,
\]
is the mean energy of the ground and first excited state of the system in the united atom
(u.a.) limit, respectively. The analytic approximation for mean energy $E_0$ which mimics
the asymptotic expansions for small \re{VRsmall} and large \re{VRlarge} distances is again
two-point Pad\'e approximation of the form~\re{Pade}
\begin{equation}
     E_0(R)_{\{n_0,n_{\infty}\}}= \frac{Z^2}{R}\ \frac{P_{N}(R)}{Q_{N+3}(R)}\bigg|_{\{n_0,n_{\infty}\}}
     \equiv
     \frac{1}{R}\ {\rm Pade}[N/N+3]_{\{n_0,n_{\infty}\}}(R)\ .
\end{equation}
This approximation supposes to reproduce the first $n_0$ terms at small and the first $n_{\infty}$ terms at
large internuclear distances expansion for mean energy $E_{u.a.}$.

\bigskip
This procedure has already been applied successfully to the diatomic molecular hydrogen ion H$_2^+$ $(p,p,e)$~\cite{OT:2018}. In order to illustrate further an approach to the general theory of the potential curves presented above the diatomic molecular ion He$_2^+$  $(\al,\al,3e)$ will be considered as an example. Our concrete goal is to construct a simple analytic expressions for the PEC of the ground state $X^2 \Si_u^+$ and the first excited state $A^2 \Si_g^+$ in full range of internuclear distances.

\section{Molecular ion H\lowercase{e}${}_2^+$ as the example}

\subsection{Introduction}

Theoretical studies of He$_2^+$ have been carried out for many years since the pioneering
work by L.~Pauling~\cite{LP:1933}. Already in that paper it was found out explicitly that the ground state PEC exhibits a well-pronounced minimum indicating the existence of the molecular
ion He$_2^+$. This observation by Pauling was confirmed later in subsequent theoretical studies (see e.g.~\cite{AH:1991,XPG:2005} and references therein). However, despite the fast development of numerical methods and computer power an accurate description of the potential energy curves for the He$_2^+$ is still running. Only recently, the PEC for the ground state was presented with absolute accuracy
of $0.05$ cm$^{-1}$ in domain $R \in [0.9, 100.]$\,a.u. in a form of mesh of the size
0.1\,a.u. for small and 1\,a.u. for large internuclear distances, minimum of the potential well
was localized at $R_{eq}=2.042$~a.u. ~\cite{TPA-HELIUM:2012}.
The same time the present authors are unaware about studies for $R < 0.9$\,a.u.
It was found the ground state electronic term for $R > 0.9$\,a.u. is very smooth curve
without any irregularities. Above-mentioned accuracy but for excited states has not been
yet achieved to the best of the present author's knowledge. Note that for the first excited
state $A^2 \Si_g^+$ it was found some irregularity in PEC at distances smaller than
equilibrium one, $R < R_{eq}$ (see \cite{AH:1991} and references therein),
{  probably,} due to level quasi-crossing(s) with higher excited states of
the same symmetry, meanwhile the Van-der-Waals minimum occurs at large distances $R_{eq} \sim 9$\,a.u.
The influence of that irregularity of the potential curve to rovibrational spectra needs
to be investigated.
Finally, the knowledge of accurate analytic expressions for the PEC allows us to calculate
easily the rotational and vibrational states by solving the Schr\"odinger equation for the
nuclear motion with analytic potential.

\subsection{The Energy Gap $\Delta E$}

Let us start considering the behavior of the energy gap $\De E$ between the excited
state $A^2 \Si_g^+$  and the ground state $X^2 \Si_u^+$,
$$\De E=E_{A^2 \Si_g^+}-E_{X^2 \Si_u^+}\ .$$
Following Bingel \cite{WAB:1959} for small internuclear distances $R\rightarrow 0$, the
behavior is given by
\begin{equation}
\label{eq10}
\Delta E =  \de_0 + 0\cdot R + O(R^2)\ ,
\end{equation}
where
\begin{displaymath}
\delta_0=E^{{\rm Be}^+}_{2^1P_{1/2}}- E^{{\rm Be}^+}_{2^1S_{1/2}}\ ,
\end{displaymath}
is the difference between the (rounded) energies of the Beryllium ion Be${}^+$~\cite{PP:2008},
\begin{eqnarray}
\label{Ebely}
E^{{\rm Be}^+}_{2^1S_{1/2}} & = &  -28.649\,526\,{\rm Ry}\ , \\
E^{{\rm Be}^+}_{2^1P_{1/2}} & = &  -28.358\,666\,{\rm Ry}\ .\non
\end{eqnarray}
As for large internuclear distances $R\rightarrow \infty$, the  energy gap $\Delta E$ is given by~\cite{Chang:1995,S:2001}
\begin{equation}
\label{eq11}
\Delta E =  R^{1/2} e^{-\alpha_0 R}\left[\epsilon_0+\frac{\epsilon_1}{R} +\frac{\epsilon_2}{R^2}+\cdots\right]\ ,
\end{equation}
where $\alpha_0 = 1.344$, $\epsilon_0=6.608\,573$, $\epsilon_1=2.296\,763$ and $\epsilon_2=0.252\,798$.
Now, we look for an expression that interpolates the expansions \re{eq10} and \re{eq11}.
In order to do that, a new variable is introduced
\begin{displaymath}
r=\sqrt{R}\ ,
\end{displaymath}
and at the same time the parameters $\epsilon_1$ and $\epsilon_2$ are released,
which gives more flexibility to the approximation. The two-point Pad\'e-type approximation
is given by
\begin{equation}
    e^{-1.344\, r^2}\, \mbox{Pade}[N+1/N](r)\ .
\end{equation}
Explicitly, taking $N=11$,
\begin{equation}
\label{eq13}
\Delta E_{\{2,1\}}\ =\ e^{-\al_0 r^2}\frac{\de_0\ +\ \al_0 \de_0r^2\ +\ \sum_{i=2}^{5}
a_i r^{2i}\ +\ \epsilon_0 r^{12}}{1+b_1 r^7\ +\ b_2r^9\ +\ r^{11}}\ .
\end{equation}
After making fit with the numerical results of~\cite{XPG:2005} the six free parameters
take values
\begin{eqnarray*}
a_2 = -123.748\ , &&   b_1 = -1.15654\ ,\\
a_3 = \phm  214.186\ , &&   b_2 = \phm  4.04014\ ,\\
a_4 = -108.275\ , &&  \\
a_5 =  \phm 46.8906\ . &&
\end{eqnarray*}
The asymptotic behavior of the expression for $\Delta E_{\{2,1\}}$~\re{eq13} reproduces
exactly the first two terms for small internuclear distances $R\rightarrow 0$ \re{eq10}, $n_0=2$, and one term for large internuclear distances $R\rightarrow \infty$ \re{eq11}, $n_{\infty}=1$.

\begin{figure}[h!]
\begin{center}
\includegraphics[width=10cm]{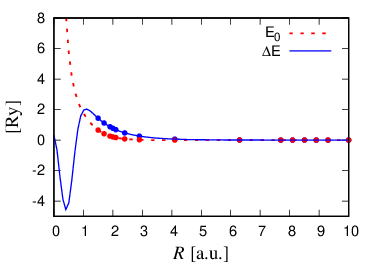}
\caption{Fits for the mean energy $E_0$ (dotted line) from \re{eq7} and
the energy gap $\De E$ (solid line) from \re{eq13}. Points represent
the numerical results~\cite{XPG:2005}, where available.}
\label{fe0ed}
\end{center}
\end{figure}

Comparison between the fit~\re{eq13} and the numerical results~\cite{XPG:2005} is
presented in Fig.~\ref{fe0ed}.

\subsection{The mean energy $E_0$}

The dissociation energy for the ground $X^2 \Si_u^+$ and the first excited state
$A^2 \Si_g^+$ at small internuclear distances $R\rightarrow 0$ is given by
\begin{eqnarray}
\label{eq1}
   \tilde E_{X^2 \Si_u^+}^{(0)} &=& \frac{2Z^2}{R} +
(E^{{\rm Be}^+}_{2^1S_{1/2}} + |E_{\infty}|)+ 0\cdot R + O(R^2)\non \ ,\\
    \tilde E_{A^2 \Si_g^+}^{(0)} &=& \frac{2Z^2}{R} +
(E^{{\rm Be}^+}_{2^1P_{1/2}} + |E_{\infty}|)+ 0\cdot R + O(R^2) \ ,
\end{eqnarray}
where $Z=2$, $E^{{\rm Be}^+}_{2^1S_{1/2}}$ and $E^{{\rm Be}^+}_{2^1P_{1/2}}$  are given
by~\re{Ebely} and
\[
E_{\infty}= E_{\rm He} + E_{{\rm He^+}} =-(5.807\,449\,+\, 4.000\,000)\, {\rm Ry}=-9.807\,449\,{\rm Ry}\ ,
\]
is the asymptotic energy of the diatomic molecule He$_2^+$.

The mean energy $E_0$
\begin{equation}
E_0 = \frac{\tilde E_{X^2 \Sigma_u^+}^{(0)} +\tilde E_{A^2 \Sigma_g^+}^{(0)}}{2}\ ,
\end{equation}
at small internuclear distances $R\rightarrow 0$, see \re{eq1},
\begin{equation}
\label{eq5}
E_0 = \frac{2Z^2}{R} + C_0 + 0\cdot R + O(R^2)\ ,
\end{equation}
where $C_0  = (E^{{\rm Be}^+}_{S}+E^{{\rm Be}^+}_{P}\,+\,2\,|E_{\infty}|)/2$.

On the other hand, at large internuclear distances $R\rightarrow \infty$, the expansion of $E_0$, obtained from the asymptotic expressions of the energy of the ground and first excited
states,
\begin{equation}
\label{erinf}
\tilde E_{X^2 \Si_u^+ / A^2 \Si_g^+}^{(\infty)}\ =\  - \frac{C_4}{R^4}\
-\ \frac{C_6}{R^6}\  +\ \cdots\ \mp\ \frac{1}{2} \De E\ ,
\end{equation}
has the form
\begin{equation}
\label{eq6}
E_{0}\ =\  - \frac{C_4}{R^4}\ -\ \frac{C_6}{R^6}\ +\ \cdots\ ,
\end{equation}
with
\begin{eqnarray*}
C_4 &\ =\ & 1.382874\ ,\\
C_6 &\ =\ & 3.193540\ ,
\end{eqnarray*}
see \cite{XPG:2005}.

In order to interpolate the two asymptotic expansions~\re{eq5} and \re{eq6} we use the two-point Pad\'e approximant
\[
 \frac{1}{R}\,\mbox{Pade}[N/N+3]_{\{3,3\}}(R)\ ,
\]
where it is required that the first three terms of the expansions at both small and large distances should be reproduced exactly. Explicitly, by choosing the approximant at $N=6$ we get
\begin{equation}
\label{eq7}
   E_{0 _{\{3,3\}}}\ =\ \frac{8+\sum_{i=1}^{5} a_iR^{i}-C_4R^6
}{R(1+\al_1\,R+\al_2\,R^2+\sum_{i=3}^{6} b_i\,R^{i}-\al_7\,R^7-\al_8\,R^8+R^9)}\ ,
\end{equation}
with four constraints between parameters
\begin{eqnarray}
\al_1 & = & (a_1-C_0)/8\ ,\\
\al_2 & = & (8 a_2-a_1 C_0 + C_0^2)/64\ ,\non \\
\al_7 & = & (C_6+a_4)/C_4\ ,\non \\
\al_8 & = & a_5/C_4\ .\non
\end{eqnarray}
These constraints guarantee the first three terms in each expansion~\re{eq5} and \re{eq6}
are reproduced exactly when (\ref{eq7}) is expanded at small and large $R$. The remaining nine parameters in (\ref{eq7}) are fixed fitting the all numerical results obtained in~\cite{XPG:2005}
\begin{eqnarray*}
a_1=& 471.867\ ,  & b_3= 103.213\ ,\\
a_2=&-706.524\ ,  & b_4=-515.786\ ,\\
a_3=& 474.091\ ,  & b_5= 623.091\ ,\\
a_4=&-148.695\ ,  & b_6=-350.333\ ,\\
a_5=&21.8549\ .   &
\end{eqnarray*}
Comparison of the fit~\re{eq7} and the numerical results~\cite{XPG:2005} is done on Fig.~\ref{fe0ed}.

\subsection{Potential Energy Curves}

Explicit analytic expressions for mean energy $E_0$~\re{eq7} and energy gap
$\De E$~\re{eq13} allow us to recover the potential energy curves for the ground
$X^2 \Sigma_u^+$ and first excited $A^2 \Si_g^+$  states,
\begin{equation}
\label{eq14}
  E_{X^2 \Si_u^+ / A^2 \Si_g^+} = E_0 \mp \frac{1}{2}\De E \quad .
\end{equation}
As a result it reproduces with accuracy 3-4 s.d. the total energy for the ground
state and for the first excited state in the entire domain in $R > 0$ when compared with
results \cite{XPG:2005}, it is shown in Table~\ref{tt1}, with exception of the domain
$0.5 \leq R \leq 1.5$\,a.u. for the first excited state $A^2 \Si_g^+$,
where the deviation is significant.

The minimum of the ground state $X^2 \Si_u^+$ electronic term calculated by vanishing
the first derivative of the expression~\re{eq14} gives $E_t = -0.181\,64$~Ry at
$R_{eq}=2.041$~a.u. For comparison, the {\it ab initio} calculations carried out
in~\cite{XPG:2005} give $E_t = -0.181\,76$~Ry at $R_{eq}=2.043$~a.u.\,,
while the most accurate numerical result at present leads to $E_t = -0.181\,84$~Ry
at $R_{eq}=2.042$~a.u.~\cite{TPA-HELIUM:2012}. The crossing of the potential curve~\re{eq14}
with the horizontal line $E=0$ occurs at $R_0=1.4083$\,a.u. in agreement
with~\cite{XPG:2005}, where it should be inside the domain $1.4 < R_0 < 1.5$\,a.u.

The Van-der-Waals minimum for the state $A^2 \Si_g^+$ is located at
$R=8.741$~a.u. with depth  $E_t = -0.000\,158$~Ry~\cite{XPG:2005},
while ~\re{eq14} predicts $R = 8.362$~a.u. with $E_t=-0.000\,198$~Ry.
Note that the fit ~\re{eq14} predicts the crossing point $R_0=7.1066$\,a.u.
in agreement with the results by \cite{XPG:2005}: $6.7 < R_0 < 7.4$\,a.u.

\begin{figure}[h!]
\begin{center}
\includegraphics[scale=1.5]{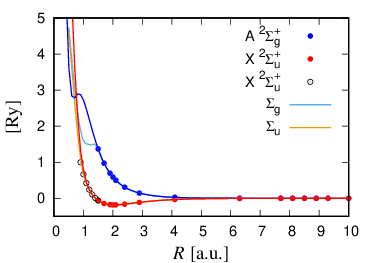}
\caption{Potential energy curves obtained from~\re{eq14} (marked by red and dark blue)
 compared with numerical results (marked by dots) from \cite{XPG:2005} (red and dark blue),
 \cite{TPA-HELIUM:2012} (empty circles).
 Curves indicated as $\Si_{g,u}$ (light blue and yellow) taken from~\cite{AH:1991}.}
\label{fpotC}
\end{center}
\end{figure}

As it is seen in Table~\ref{tt1} even though the simple analytic approximation~\re{eq14}
predicts reasonably correct the position and the depth of the minima for both states,
comparison with results by~\cite{AH:1991} reveals at $R <1.5$~a.u. a significant deviation
for the excited state $A^2 \Si_g^+$  as well as a small insignificant deviation
for the ground state $X^2 \Si_u^+$, see~\cite{TPA-HELIUM:2012}. For the state $A^2 \Si_g^+$
the PEC displays irregularity at $0.5 \lesssim R \lesssim 1.5$\,a.u. which can be attributed
to a quasi-crossing with the next $\Si_g$ excited state. Interestingly, the pattern of
irregularity presented in~\cite{AH:1991} and in (\ref{eq14}) are similar qualitatively,
see Fig.~\ref{fpotC}. Note that the irregularity occurs for energies much above the threshold
energy $E(He) + E(He^+)$, it should not bring much influence to the rovibrational spectra.

Surprisingly, for $X^2 \Si_u^+$ state fit \re{eq14} also predicts a certain irregularity
in the domain of $0.9 \lesssim R \lesssim 1.5$\,a.u.  as it can be seen in
Fig.~\ref{fpotC}: Numerical data from \cite{TPA-HELIUM:2012} deviate from our analytic curve
as well as from one obtained in \cite{AH:1991} in this domain.
Since it is relatively small and is situated far above the threshold energy we do not expect
much influence to the rovibrational spectra. Subsequent calculations confirm this prediction,
see below.

It can be checked that the asymptotic expansion of fit~\re{eq14} for the ground state
$X^2 \Si_u^+$ at $R \rar 0$ is given by
\begin{equation}
\label{eq15-1}
    E_{X^2 \Si_u^+ } \ =\  \frac{8}{R}\ -\ 18.842078\ +\ 0 \cdot R\ +\ \cdots\\
\end{equation}
while the expansion at $R \rar \infty$ has the form
\begin{equation}
\label{eq15-2}
  E_{X^2 \Si_u^+ }\ =\ -\ \frac{1.382874}{R^4}\ -\ \frac{3.193540}{R^6}\ +\ \ldots
  \ - \ e^{-1.344\,R}\,R^{1/2}\left[3.304287\ +\ \frac{10.095520}{R}\ +\ \cdots\right]
\ .
\end{equation}
As for the excited state $A^2 \Si_g^+$ the asymptotic behavior of~\re{eq14} are
\begin{eqnarray}
\label{eq16}
   E_{A^2 \Si_g^+ } &=& \frac{8}{R}\ -\ 18.551218\ +\ 0\cdot R\ +\ \cdots\\
    E_{A^2 \Si_g^+ } &=& \ -\frac{1.382874}{R^4}\ -\ \frac{3.193540}{R^6} \cdots
    +\ e^{-1.344\,R}\,R^{1/2}\left[3.304287\ +\ \frac{10.095520}{R}\ +\ \cdots\right]
\ .\non
\end{eqnarray}
For both states these expansions are in agreement with the asymptotic behavior (cf.~\re{eq1} and~\re{erinf}).

\begin{table}
\caption{Energy of the ground $X ^2\Si_u^+$ and the excited state $A ^2\Si_g^+$ of
the molecular ion He$_2^+$ obtained using approximation~\re{eq14}. The second and third
lines for given $R$ display the results of~\cite{XPG:2005} and~\cite{TPA-HELIUM:2012}, respectively. For $R=1.0$\,a.u. the second line result is from~\cite{AH:1991}.}
\begin{center}
\resizebox{6.0cm}{!}{
\begin{tabular}{lll | cll}
\hline\hline
$R$ & $X ^2\Si_u^+$ & $A ^2\Si_g^+$&$R$&$X ^2\Si_u^+$&$A ^2\Sigma_g^+$\\
\hline
1.0 \hspace{0.2cm} &  0.78489   & 2.72578  & \hspace{0.1cm} 2.65 \hspace{0.2cm} & -0.13865   & 0.21102  \\
  &  0.66628   & 1.59046  &     & -0.13846   & 0.210960 \\
  &  0.66537   &          &     &            &          \\
1.1  &  0.44602   &          &2.9  & -0.11365   & 0.14549  \\
  &  0.42043   &          &     & -0.11358   & 0.145398 \\
  &  0.42014   &          &     & -0.11360   &          \\
1.2  &  0.23859   &          &3.5  & -0.06425   & 0.06189  \\
  &  0.23910   &          &     & -0.06454   & 0.061922 \\
  &  0.23884   &          &     & -0.06456   &          \\
1.3  &  0.10222   &          &4.1  & -0.03416   & 0.02682  \\
  &  0.10544   &          &     & -0.0342    & 0.027232 \\
  &  0.10521   &          &     & -0.03420   &          \\
1.4  &  0.00662   &          &5.3  & -0.00927   & 0.00462  \\
  &  0.00787   &          &     & -0.00892   & 0.003852 \\
  &  0.00766   &          &     & -0.00891   &          \\
1.5  & -0.06220   & 1.36915  &6.3  & -0.00319   & 0.00077  \\
  & -0.06218   & 1.36908  &     & -0.00296   & 0.00072  \\
  & -0.06236   &          &     & -0.00296   &          \\
1.7  & -0.14419   & 0.97461  &6.9  & -0.00174   & 0.00011  \\
  & -0.14416   & 0.974356 &     & -0.0016    & 0.000210 \\
  & -0.14431   &          &     & -0.00159   &          \\
1.8  & -0.16507   & 0.82239  &9.3  & -0.00025   &-0.00017  \\
  & -0.16496   & 0.822274 &     & -0.00024   &-0.000148 \\
  & -0.16508   &          &     & -0.00023   &          \\
1.9  & -0.17662   & 0.69514  &9.6  & -0.00021   &-0.00015  \\
  & -0.17656   & 0.695202 &     & -0.000198  &-0.00014  \\
  & -0.17668   &          &     & -0.000198  &          \\
2.0  & -0.18126   & 0.58884  &10.0 & -0.00017   &-0.00014  \\
  & -0.18132   & 0.588948 &     & -0.00016   &-0.000126 \\
  & -0.18143   &          &     & -0.000160  &          \\
2.1  & -0.18092   & 0.49991  &10.5 & -0.00013   &-0.00012  \\
  & -0.18106   & 0.499976 &     & -0.000126  &-0.000108 \\
  & -0.18115   &          &     & -0.000126  &          \\
2.2  & -0.17703   & 0.42534  &11.0 & -0.00011   &-0.000098 \\
  & -0.17716   & 0.425350 &     & -0.000102  &-0.000092 \\
  & -0.17723   &          &     & -0.0001015 &          \\
2.4  & -0.16260   & 0.30989  &12.0 & -0.00007   &-0.00007  \\
  & -0.16254   & 0.309830 &     & -0.00007   &-0.000066 \\
  & -0.16260   &          &     & -0.000069  &          \\
\hline\hline
\end{tabular}}
\end{center}
\label{tt1}
\end{table}

\subsection{Rovibrational States}

In the Born-Oppenheimer approximation the rovibrational states are calculated
by solving the one-dimensional radial Schr\"odinger equation for the nuclear motion
\begin{equation}
\label{eq31}
\left[-\frac{1}{\mu}\frac{d^2}{dR^2}\ +\ \frac{L(L+1)}{\mu R^2}\ +\ V(R)\right]\,\phi(R)\
=\  E_{\nu,L}\, \phi(R)\ ,
\end{equation}
where $$\mu = M_{\al}/2=3647.149771\ldots\ ,$$ is the reduced mass of two $\al$
particles, here $\nu$ and $L$ are the vibrational and rotational quantum numbers,
respectively: any state will be marked as ($\nu,L$). It implies that we study the
rovibrational state of the diatomic ion ${}^4$He$_2^{+}$. Usually, the
equation~\re{eq31} is solved numerically with the potential $V(R)$ defined at
discrete set of points in $R$ as the result of the {\it ab initio} calculations. In our
case the potential $V(R)$ is given by some analytic expression for the potential
energy curve~\re{eq14} (emerging from the expressions \re{eq13} and \re{eq7}).
In this case the Lagrange-mesh method~\cite{DB:2015} can be used in its generality
in a very efficient and economic way.

\begin{figure}[h!]
\begin{center}
\includegraphics[scale=1.5]{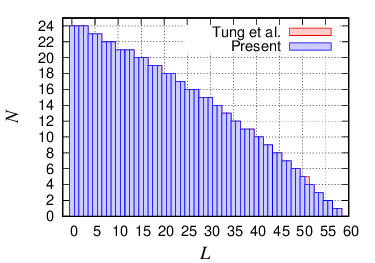}
\caption{$^4$He${}_2^+$ molecular ion: the histogram of rovibrational states
supported by the ground state $X^2\Si_u^+$ as a function of the angular
momentum $L$ with $L_{max} = 58$. In total, there are 829 rovibrational
states marked by blue in~\cite{TPA-HELIUM:2012} and confirmed in present paper.
The extra, weakly-bound state, marked by red, is reported in~\cite{TPA-HELIUM:2012}, where
adiabatic corrections are included unlike the present work.}
\label{hepH}
\end{center}
\end{figure}

It can be immediately obtained using the Lagrange mesh method that the PEC for
the ground state $X^2 \Si_u^+$ \re{eq14} supports 24 vibrational states $(\nu,0)$
and 59 rotational states $(0,L)$, thus, with $L_{max} = 58$. In total, we found 829 rovibrational
states $(\nu,L)$, one state less than the 830 states reported previously
in~\cite{TPA-HELIUM:2012}. It is worth mentioning that the rovibrational state energies
in \cite{TPA-HELIUM:2012} were obtained with including adiabatic corrections into the PEC.
All found states are presented in a histogram in Fig.~\ref{hepH}. Making a careful comparison of our results with those in~\cite{TPA-HELIUM:2012} one can see that they agree in 3-4 s.d.

In BO approximation the PEC for ${}^3$He$_2^{+}$ and ${}^4$He$_2^{+}$ coincide by construction, however, small adiabatic
corrections make these curves different, usually, in the 4th s.d., see for discussion \cite{TPA-HELIUM:2012}. It is natural
to assume that rovibrational energies are changed by adiabatic corrections in 3-4 s.d., which is consistent with accuracy
of analytic approximations of the PEC we use in present work.
With the analytic expression for the PEC~\re{eq14} in BO approximation, by making a simple replacement of the nuclear mass
of ${}^4$He ($\al$ particle) by the nuclear mass of ${}^3$He (in the reduced mass) in the radial Schr\"odinger equation
\re{eq31}, the rovibrational spectra for diatomics ${}^3$He$_2^{+}$ can be recalculated.
Considering $$m_{{}^3 {\textrm He}}=5495.88512\,a.u.\ ,$$ see~\cite{AWT:2003}, one can find that the ground state keeps
21 vibrational states $E_{(\nu,0)}$, $\nu=0,\cdots,20$ in agreement with~\cite{TPA-HELIUM:2012} in 3-4 s.d. ,
while the maximum angular momentum is $L_{max}=50$.
In total, the state  $X^2\Si_u^+$ supports  626 rovibrational states, one state more
than the 625 states reported in~\cite{TPA-HELIUM:2012}. Note that with decrease in reduced mass
of two-body system both the number of (ro)vibrational states and maximal angular momentum get smaller.

Applying the same procedure for the PEC of the first excited state  $A ^2\Si_g^+$, our
results point out the presence of nine rovibrational states, the same number of
states as was found in~\cite{XPG:2005} (see Table~\ref{rovibEx}). In general, the obtained rovibrational energies
are very small being of the order $10^{-5} - 10^{-6}$\,Ry
(or even less), while one can speculate that adiabatic corrections (as well as relativistic and QED corrections)
are of the order $1/\mu \sim 10^{-4}$ as well as the accuracy with PEC was obtained. Even though all our results
are stable inside of the Lagrange-mesh method, they fall well beyond our precision. It is worth mentioning that
we predict the rotational state $(0,5)$, which is not found in \cite{XPG:2005}, while in \cite{XPG:2005} it is predicted
the vibrational state $(2,0)$ with the energy of the order of $10^{-9}$\,Ry, which is not seen in our calculations.
It is clear that this state should be discarded.

\begin{table}
\caption{Molecular ion $^4$He$_2^+$: Rovibrational energies $E_{(\nu,L)} \times 10^{-5}$~Ry
        for the state $A ^2\Si_g^+$ in approximation~\re{eq14}.
        The second line displays results from~\cite{XPG:2005}.}
\begin{center}
\begin{tabular}{c|lll}
\hline \hline
    \ $L$ \ &\  $\nu=0$ \ &\ $\nu=1$ \ &\ $\nu=2$
\\
\hline
  0\ &\  -9.74 \   &\  -1.17 \   &\     \\
     &\  -7.3762\  &\  -0.7194 \ &\ -0.0003 \\
  1\ &\  -9.12 \   &\  -0.90 \   &\     \\
     &\  -6.8212\  &\  -0.5074 \ &\     \\
  2\ &\  -7.90 \   &\  -0.42 \   &\     \\
     &\  -5.7289 \ &\  -0.1357 \ &\     \\
  3\ &\  -6.11 \   &\            &\     \\
     &\  -4.1381 \ &\            &\     \\
  4\ &\  -3.81 \   &\            &\     \\
     &\  -2.1200 \ &\            &\     \\
  5\ &\  -1.11 \   &\            &\     \\
\hline\hline
\end{tabular}
\end{center}
\label{rovibEx}
\end{table}

\subsection{He${}_2^+$: Conclusions}

Inside of the Born-Oppenheimer approximation by using two-point Pad\'e approximants,
analytic expressions for the potential energy curves in whole range of internuclear
distances $R > 0$ are constructed for both the ground  $X^2\Si_u^+$ and the first excited
$A ^2\Si_g^+$ states for the diatomic molecular ion He$_2^+$.  In general, the obtained
analytic curves reproduce known numerical results with an accuracy of 3-4 s.d.
in the whole domain in $R$.

For small internuclear distances $0.5 < R < 1.5$~a.u., possibly due to the
quasi-crossing
between the excited state $A ^2\Si_g^+$ with the next $\Si_g$ excited states the potential
energy curve gets inaccurate in this domain of internuclear distances. It leads to a
certain loss of accuracy in the spectra of rovibrational states situated in Van-der-Waals
minimum, but it does not change the number of rovibrational states, which is equal to
nine. All these states are very weakly-bound.

In the case of ground state $X^2\Si_u^+$ the predicted potential curve through analytic
approximation \re{eq14} (with the expressions \re{eq13} and \re{eq7} as
building blocks) in the same domain $0.9 < R < 1.5$~a.u. differs from numerical results
insignificantly \cite{TPA-HELIUM:2012}. Probably, it indicates to the existence of ``weak"
quasi-crossing effect. This deviation does not make significant change in description
of the spectra of rovibrational states obtained with 3-4 s.d. in accuracy.
Note the predicted minima by the approximations \re{eq14} differ from the numerical
results by $\sim 0.1\%$ and $\sim 25\%$ for the ground $X^2\Si_u^+$ and the first excited
$A^2\Si_g^+$, respectively.

The obtained analytic expressions for the PEC allow us to solve the radial (nuclear)
Schr\"odinger equation for the nuclear motion of ${}^4$He$_2^{+}$ by using the Lagrange-mesh
method with an accuracy of 3-4 s.d. The ground state curve $X^2\Si_u^+$ keeps 829 rovibrational
states, which is one state less than the 830 states reported in the literature~\cite{TPA-HELIUM:2012},
where adiabatic corrections are taken into account. For the excited state curve $A ^2\Si_g^+$
the predicted rotational and vibrational states are certainly beyond of the BO approximation
and various corrections should be taken into account in order to make definite conclusions.
By replacing the nucleus ${}^4$He by nucleus ${}^3$He in the radial (nuclear) Schr\"odinger
equation (\ref{eq31}), the rovibrational spectra for diatomics ${}^3$He$_2^{+}$
can be calculated. As expected, the number of rovibrational states is significantly
reduced.

The calculated rovibrational states $(\nu,L)$ due to the analytic knowledge of PEC \re{eq14},
allow us to explore radiative transitions between those states as it was done in~\cite{OB:2012,OT:2016}.
Up to our knowledge, radiative transitions for the molecular ion He$_2^+$ have not been considered before.
It will be done elsewhere.

\section{Molecular ion L\lowercase{i}H as the example}

\begin{figure}[h!]
\begin{center}
\includegraphics[width=10cm]{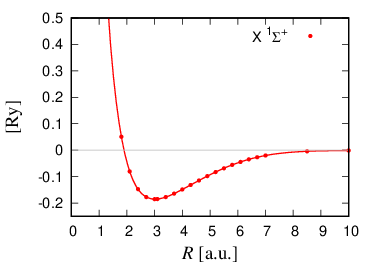}
\caption{PEC for the diatomic molecule LiH from~\re{LiHV} (solid line).
         Points represent data from~\cite{TPA-LITHIUM:2011}.}
\label{potLiH}
\end{center}
\end{figure}

As an example of the application of our approach to heteronuclear diatomic molecules,
let us consider the ground state $X^1 \Si^+$ of the LiH molecule.  At small
internuclear distances $R\rightarrow 0$ the dissociation energy $\tilde{E}$ is given by
\begin{equation}
\label{lihR0}
\tilde E_{X^1 \Si^+} = \frac{2\,Z_1\,Z_2}{R} +\ep_0+ 0\cdot R + O(R^2) \ ,
\end{equation}
where $Z_1=3\,,\, Z_2=1$, and $\ep_0= E^{\rm Be} + |E_{\infty}|$. The energy of the united atom is
$E^{{\rm Be}} =-29.33474$~Ry~\cite{MAA:1991}   and
\begin{displaymath}
E_{\infty}= E_{\rm H} + E_{{\rm Li}} =-(1.000\,000\,+\, 14.956\,120)\, {\rm Ry}=-15.956\,120\,{\rm Ry}\ ,
\end{displaymath}
is the so-called asymptotic energy~\cite{PP:2006}.
For large internuclear distances $R\rightarrow \infty$, the  energy is given by~\cite{JMCB:2015}
\begin{equation}
\label{lihRl}
\tilde E_{X^1 \Si^+}  =  -\frac{c_6}{R^6} -\frac{c_8}{R^8}+\cdots,
\end{equation}
where $c_6 = 133.182$ is the Van-der-Waals constant. Let us take two-point Pad\'e approximant given by
\begin{equation*}
    \frac{1}{R}\, \mbox{Pade}[N/N+5](R)\ ,
\end{equation*}
see (\ref{PadeN}), choose $N=4$,
\begin{equation}
\label{LiHV}
   \tilde E_{X^1 \Si^+\, _{\{3,2\}}}  = \frac{6+a_1 R+ a_2 R^2+a_3 R^3+c_6 R^4}
   {R\,(1+\al_1 R+\al_2 R^2+\sum_{i=3}^{7}b_i R^i +\al_3 R^8+R^9)}\ ,
\end{equation}
and impose three constraints
\begin{eqnarray}
\label{paramLiHV-3}
\al_1&=&(a_1-\ep_0)/6\ ,\\
\al_2&=&(6 a_2-a_1 \ep_0+\ep_0^2)/36\ ,\non\\
\al_3&=&a_3/c_6\ ,\non
\end{eqnarray}
which guarantee that the expansions of $\tilde E_{X^1 \Si^+\, _{\{3,2\}}}$~\re{LiHV} at small and large internuclear distances reproduces correctly the first three terms for small internuclear distances, $R^{-1}$, $R^0$ and $R^1$ in~\re{lihR0} and the first
two terms for large internuclear distances, $R^{-6}$ and $R^{-7}$,  in~\re{lihRl}.
Making fit of data~\cite{TPA-LITHIUM:2011} with ~\re{LiHV} or choosing eight points in {\it ab initio} calculations, which look confident, we find eight free parameters
\begin{eqnarray*}
 a_1\ =\ \phm  44744.6\ , &&\ b_4\ =\  \phm  2179.81\ ,\\
 a_2\ =\ -28086.2\ ,      &&\ b_5\ =\  \phm  727.108\ ,\\
 a_3\ =\ \phm  2614.80\ , &&\ b_6\ =\ -592.941\ ,\\
 b_3\ =\ -6739.29\ ,      &&\ b_7\ =\  \phm  158.198\ .
\label{paramLiHV-8}
\end{eqnarray*}
Comparison of fit~\re{LiHV} with parameters (\ref{paramLiHV-3}), (\ref{paramLiHV-8}) with the numerical results from~\cite{TPA-LITHIUM:2011} is presented in Table~\ref{ttELiH} and illustrated by Fig.~\ref{potLiH}.
Note that fit reproduces with high
accuracy the equilibrium distance $R_e=3.015$\,a.u., see e.g.~\cite{TPA-LITHIUM:2011}, without extra tuning, and also its vicinity.
Fit predicts that the potential curve vanishes, $E(R_0)=0$, at $R_0=1.8954$\,a.u., where the dissociation energy is equal
to zero. For larger $R > R_0$ PEC becomes negative.

\begin{table}
\caption{ LiH molecule: PEC for the ground state $X^1 \Si^+$ the obtained
using approximation~\re{LiHV}. Third column displays the results of~\cite{TPA-LITHIUM:2011}
(rounded).}
\begin{center}
\resizebox{3.2cm}{!}{
\begin{tabular}{lrr}
\hline\hline
$R$  &\ fit\re{LiHV}\ &\ \cite{TPA-LITHIUM:2011} \\
\hline
 1.8   &  0.05099\ &\  0.05100\\
 1.9   & -0.00221\ &\ -0.00221\\
 2.0   & -0.04539\ &\ -0.04540\\
 2.2   & -0.10809\ &\ -0.10810\\
 2.4   & -0.14732\ &\ -0.14730\\
 2.6   & -0.17013\ &\ -0.17011\\
 2.8   & -0.18151\ &\ -0.18151\\
 3.0   & -0.18495\ &\ -0.18496\\
 3.015 & -0.18497\ &\ -0.18497\\
 3.1   & -0.18450 & -0.18451\\
 3.2   & -0.18293 & -0.18294\\
 3.4   & -0.17719 & -0.17719\\
 3.6   & -0.16897 & -0.16896\\
 3.8   & -0.15916 & -0.15915\\
 4.0   & -0.14840 & -0.14838\\
 4.2   & -0.13714 & -0.13713\\
 4.4   & -0.12573 & -0.12572\\
 4.6   & -0.11440 & -0.11441\\
 4.8   & -0.10335 & -0.10335\\
 5.0   & -0.09269 & -0.09269\\
 5.5   & -0.06832 & -0.06831\\
 6.0   & -0.04784 & -0.04783\\
6.5   & -0.03172 & -0.03171\\
 7.0   & -0.01997 & -0.01998\\
 7.5   & -0.01209 & -0.01211\\
 8.0   & -0.00716 & -0.00719\\
 8.5   & -0.00422 & -0.00424\\
 9.0   & -0.00251 & -0.00250\\
 9.5   & -0.00152 & -0.00148\\
10.0   & -0.00094 & -0.00089\\
11.0   & -0.00039 & -0.00033\\
12.0   & -0.00018 & -0.00013\\
13.0   & -0.00009 & -0.00006\\
14.0   & -0.00005 & -0.00003\\
\hline\hline
\end{tabular}}
\end{center}
\label{ttELiH}
\end{table}

\subsection{Rotational and vibrational states}

By solving the radial Schr\"odinger equation for the nuclear motion~\re{eq31}
with the analytic potential~\re{LiHV}, the rovibrational spectra $E_{(\nu,L)}$
can be obtained. The reduced mass of the ${}^7$LiH molecule is calculated using
$m_{{}^7 {\rm Li}} = 12786.393$ and $m_{\rm H} = 1836.153$~\cite{TPA-LITHIUM:2011}.
Following the calculations carried out in Lagrange mesh method the ground state
$X ^1\Si^+$ of the diatomics ${}^7$LiH supports 24 vibrational states
$E_{(\nu,0)}$.

Making comparison the spectra of vibrational states $E_{(\nu,0)}$ with one presented
in~\cite{TPA-LITHIUM:2011} one can see that not less than 4 figures are reproduced for $\nu \le 7$
as it can be seen in Table~\ref{tvib}. However, for $\nu \ >\ 7$ the accuracy is reduced
to 3 figures and for some values of $\nu$ even to 2 figures.
We have to note that the state ${(23,0)}$, is not reported in~\cite{TPA-LITHIUM:2011}, however,
it is identified in~\cite{DAM:2018}.

\begin{table}
\caption{Molecule $^7$LiH: Vibrational states $E_{(v,L=0)}$ in the ground state  $X ^1\Si^+$.
           For sake of convenience the 3rd column displays energies in cm$^{-1}$ with five figures
           (1 Hartree = 219\,474.631\,363 cm$^{-1}$).  4th column displays the results from in~\cite{TPA-LITHIUM:2011}.}
\begin{center}
\resizebox{6.0cm}{!}{
\begin{tabular}{rrrr}
\hline\hline
$v$  & $E_{(v,0)}$[Ry] &\ \ $E_{(v,0)}$[cm$^{-1}$]&  \cite{TPA-LITHIUM:2011}     \\
\hline
0  & -0.17861\ & -19600. &-19600.57\\
1  & -0.16621\ & -18240. &-18240.34\\
2  & -0.15423& -16925. &-16924.98\\
3  & -0.14264& -15653. &-15653.58\\
4  & -0.13145& -14425. &-14425.30\\
5  & -0.12064& -13238. &-13239.37\\
6  & -0.11021& -12094. &-12095.19\\
7  & -0.10015& -10990. &-10992.19\\
8  & -0.09047& -9927.7 & -9930.00\\
9  & -0.08116& -8905.9 & -8908.41\\
10 & -0.07222& -7924.9 & -7927.45\\
11 & -0.06365& -6984.9 & -6987.43\\
12 & -0.05546& -6086.4 & -6088.98\\
13 & -0.04767& -5230.6 & -5233.15\\
14 & -0.04027& -4419.1 & -4421.65\\
15 & -0.03330& -3654.3 & -3656.91\\
16 & -0.02679& -2939.7 & -2942.32\\
17 & -0.02078& -2279.9 & -2282.68\\
18 & -0.01532& -1681.4 & -1684.62\\
19 & -0.01050& -1152.7 & -1156.46\\
20 & -0.00643& -705.37 &  -708.64\\
21 & -0.00323& -354.58 &  -357.53\\
22 & -0.00108& -118.73 &  -118.21\\
23 & -0.00011&  -11.65 &     --  \\
\hline\hline
\end{tabular}}
\end{center}
\label{tvib}
\end{table}

\begin{figure}[h!]
\begin{center}
\includegraphics[width=10cm]{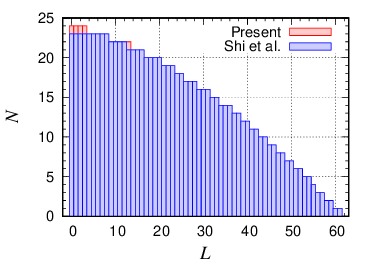}
\caption{
         Diatomic molecule $^7$LiH: 906 rovibrational bound states in the ground state
         $X ^1\Si^+$ as a function of the angular momentum $L$. Weakly bound states marked by red are the result of the present analysis (see text).}
\label{LiHhyst}
\end{center}
\end{figure}

In total, the ground state $X^1\Si^+$ of the diatomic molecule ${}^7$LiH supports 906 rovibrational states $E_{(\nu,L)}$.  These are presented by histogram in Fig.~\ref{LiHhyst} with $L_{max}=61$. In~\cite{SSW:2012}, the total of 901 rovibrational states is reported. The five extra states $E_{(23,0)}$, $E_{(23,1)}$, $E_{(23,2)}$, $E_{(23,3)}$ and $E_{(21,13)}$, which are found in the present work, are indicated by red in Fig.~\ref{LiHhyst}.
Among them, the four states $E_{(23,L)}$ with $L=0,1,2,3$ and
the state $E_{(14,35)}$ (which is not present in our calculations) are reported
in~\cite{DAM:2018} not so the state $E_{(21,13)}$.

Needless to say that the radial (nuclear) Schr\"odinger equation (\ref{eq31})
depends on the reduced mass of two body nuclear system in kinetic energy term.
As is shown by replacing the nucleus ${}^7$Li by ${}^6$Li and/or hydrogen H (proton)
by deutron or triton one can calculate the rovibrational spectra of diatomics ${}^6$LiH and ${}^{6,7}$LiD,
${}^{6,7}$LiT. Let us consider, for instance, the isotopologue ${}^6$LiH.
Taking the nuclear mass of ${}^6$Li as $m_{^6{\rm Li}}=10961.898$~\cite{AWT:2003}
it is found that there are 24 vibrational states $E_{(\nu,0)}$ bonded by the
electronic ground state $X^1\Si^+$ with $\nu=0,\cdots,23$. The maximum angular momentum is $L_{max}=60$.
In total, the  state  $X^1\Si^+$ supports 884 rovibrational states, which one state more than
the 883 reported in~\cite{DAM:2018}. In a similar way as to the homonuclear case of He$_2^+$,
with decrease of the reduced mass in heteronuclear case the number of rovibrational states and maximal angular
momentum are  reduced.

\section{Conclusions}

In studying the electronic Hamiltonian for diatomic molecule the domain of small and
large internuclear distances can be explored in sufficiently easy manner without performing
massive numerical calculations. Straightforward interpolation between these two domains by
using (generalized) two-point Pad\'e approximations, involving a few points in $R$ of order
1\,a.u. around minimum of the PEC found via {\it ab initio} calculations, leads to amazingly
accurate analytic formulas for potential curves. In this formalism hetero- $(A+B)$ and
homo-nuclear $(A+A)$ diatomics are conceptually different: latter ones contain exponentially
small terms at large distances due to tunneling between two identical Coulomb wells, that
appears in addition to multipole expansion.
It results to the general formula for the approximating the potential curve,
\begin{equation}
\label{general}
  V(R)\ =\ \frac{1}{R}\,\mbox{Pade}(n/m) (R)\ +\ \de_{Z_A,Z_B} \de_{M_A,M_B} D_0(R)
  \ e^{-S_0(R)}\mbox{Pade}(p/q) (R)\ .
\end{equation}
where $n,m=n+3(5),p,q$ are integers, see below, $Z_A,Z_B$ and $M_A,M_B$ are charges and masses of nuclei A,B,
respectively, $\de$ is the Kronecker symbol, $D_0, S_0=\al R$ as well as numbers $(p,q)$ depend on
the diatomics under investigation. In general, the parameters of the Pade approximants depend significantly
on the diatomics we study.
It excludes the existence of universal formulas for the potential curves of diatomics for
the {\it entire} domain of $R$, which is in agreement with conclusions drawn in the book
by Goodisman \cite{Goodis:1961}, but proclaimed opposite in \cite{XG:2005}.

It has to be emphasized that if we consider hetero-nuclear diatomics
the second term in (\ref{general}) disappears, we arrive at (\ref{Pade}),
\begin{equation*}
  V_{ion}(R)\ =\ \frac{Z_{A} Z_{B}}{R}\ \mbox{Pade}(N / N+3)\ ,
\end{equation*}
for positively charged diatomic and at (\ref{PadeN}),
\begin{equation*}
  V_{neutral}(R)\ =\ \frac{Z_{A} Z_{B}}{R}\ \mbox{Pade}(N / N+5)\ ,
\end{equation*}
for neutral one. These formulas look functionally similar, they can be easily used to study of
any hetero-nuclear diatomics in the whole range of internuclear distances. However, the emerging
coefficients in Pade approximants depend on the system at hand and the present authors were unable
to see hints indicating to their universal behavior, at least, so far.

The approach was illustrated by study the PECs for
$^4$He${}_2^+$,  $^3$He${}_2^+$,  ${}^7$LiH and ${}^6$LiH diatomics. With sufficiently high accuracy
the rovibrational spectra is described for both diatomics and its isotopologues.
Radiative transitions will be studied elsewhere. It has to be noted that recently in the framework
of the present formalism the overwhelming study of the PEC in the form (\ref{PadeN}) and
the rovibrational spectra was carried for HF, DF and TF molecules \cite{AG-OP:2021}.

Approximate analytic expressions for the PEC allow to simplify studies of the contribution of BO potential
(electronic) curves, which describes two atom interactions (or saying differently two-body, nucleus-nucleus
interactions), into polyatomic potential surfaces at large internuclear distances. It will be studied elsewhere.
There exists an interesting possibility to construct the approximate eigenfunctions of the nuclear Hamiltonian,
where the analytic PEC plays the role of potential. It is an open direction.

\noindent
{\bf Note added.}\ {\it After submission of this paper for publication the formula (\ref{PadeN}) is employed
to explore the ground state BO potential curve and the rovibrational spectra of the chlorine
monofluoride ClF (ArXiv: 2202.10666, pp.9 (February 2022)).
It allowed us to resolve controversy between different {\it ab initio} calculations and also with
experimental data, and for the first time to calculate the total number of vibrational, rotational,
rovibrational states.}

\section{Acknowledgements}

The research is partially supported by CONACyT grant A1-S-17364 and DGAPA grants IN113819, IN113022 (Mexico).
A.V.T. thanks PASPA-UNAM for a support during his sabbatical stay at University of Miami, where this work was finished.
H.O.P. is grateful to Instituto de Ciencias Nucleares, UNAM, Mexico for kind
hospitality, where the present study was initiated and concluded. The authors thank anonymous referee for careful reading the paper and important insights.

This work is dedicated to the memory of L.~Wolniewicz whom both authors had no privilege to meet
personally but always considered as the exemplary scientist. The authors thank J.~Karwowski
for interest to the present work and for the kind invitation to contribute to this volume.


\end{document}